\documentclass[rmp,showkeys,amssymb,groupedaddress]{revtex4}

\usepackage{graphicx} 
\usepackage{amsmath}

\begin{document}
\title{The effect of biodiversity on the Hantavirus epizootic}
\author{Ignacio D. Peixoto}
\affiliation{Instituto Balseiro, 8400 SC de Bariloche, R\'{\i}o Negro, Argentina}
\affiliation{Department of Physics and Astronomy, University of New Mexico,
Albuquerque, NM 87131, USA}
\affiliation{Consortium of the Americas for Interdisciplinary Science,
University of New Mexico, Albuquerque, NM 87131, USA}
\author{Guillermo Abramson}
\affiliation{Centro At\'{o}mico Bariloche, CONICET and Instituto
Balseiro, 8400 SC de Bariloche, R\'{\i}o Negro, Argentina}
\affiliation{Consortium of the Americas for Interdisciplinary
Science, University of New Mexico, Albuquerque, NM 87131, USA}
\email{abramson@cab.cnea.gov.ar}
\date{\today}

\begin{abstract}
We analyze a mathematical model of the epizootic of
Hantavirus in mice population, including the effect of species which
compete with the host. We show that the existence of the second
species has an important consequence for the prevalence of the
infectious agent in the host. When the two mice species survive in
the ecosystem, the competitive pressure of the second species may
lead to reduction or complete elimination of the prevalence of
infection. The transition between the disappearance of the infection
and its presence occurs at a critical value of the competitor's
population, resembling a second order phase transition in a
statistical system. The results provide a rigorous framework for the
study of the impact of biodiversity in the propagation of infectious
diseases, and further lends itself to future experimental
verification.
\end{abstract}

\keywords{biodiversity, Hantavirus, epizootic, epidemic}

\pacs{87.19.Xx, 87.23.Cc, 05.45.-a}

\maketitle

\pagestyle{myheadings} \markboth{I. D. Peixoto and G. Abramson, The effect of
biodiversity on the Hantavirus epizootic}{I. D. Peixoto and G. Abramson, The
effect of biodiversity on the Hantavirus epizootic}


\section{Introduction: The role of biodiversity}
In 1993 an outbreak of a severe disease now known as Hantavirus
Pulmonary Syndrome (HPS) struck the region of the Four Corners, in
the North American Southwest, with a mortality in excess of 50\% of
those affected. Shortly afterwards, Sin Nombre Virus (Bunyaviridae:
Hantavirus, SNV), the first Hantavirus to be discovered in the
Americas, was identified as the infectious agent responsible for
HPS. The main host of the virus was identified as one of the most
common mammals in North America, the deer mouse, \emph{Peromyscus
maniculatus}. Since then, continuous efforts in different areas of
science have been made to understand this epizootic, with the
ultimate goal of correctly assessing the risk to humans. Across all
of the Americas, new discoveries of Hantaviruses (half of them being
human pathogens and some of them responsible for a high mortality
\citep{mills98}) have led to a renewed interest in the natural
history of the host rodents.

A simple mathematical model for the spatio-temporal patterns in the
spread of this epizootic has been proposed and analyzed by the
second author and others \citep{abramson02,abramson03}. The model
takes into account several peculiarities of the Hantavirus-rodent
association. An example being, the fact that the infection does not
produce any known disease in the mice, and consequently does not
affect their death rate. The role of the spatio-temporal patterns of
the environment on the prevalence of the infection, which has been
found in field studies \citep{yates02}, has also been taken into
account. We refer the reader to \citep{abramson02,abramson03} for
details. That model was able to successfully explain several field
observations as environmentally controlled phase transitions, thus
providing an analytical support to biological hypotheses such as the
trophic cascade discussed in \citep{yates02}. Among the consequences
of the mathematical model that have correlations with the field
observations, we mention the sporadic disappearance of the infection
and the formation of ``refugia,'' whence the infection
spreads---when conditions change---in the form of waves.

\begin{figure}[ht]
\includegraphics[width=10cm]{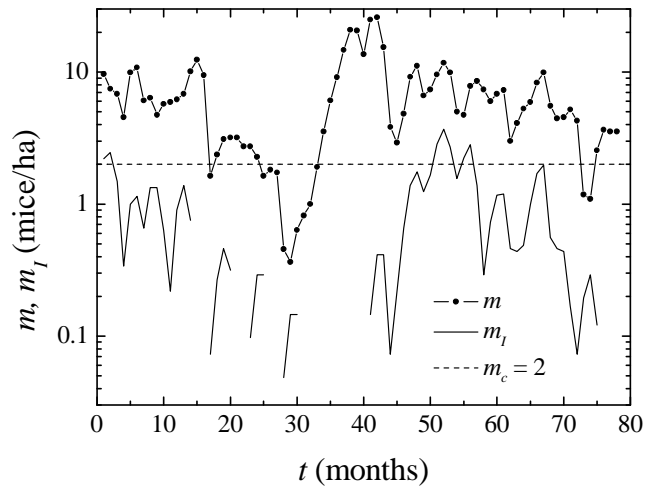}
\caption[Population threshold]{Mean density of \emph{P. maniculatus}
at two sites near Zuni, New Mexico (after \citep{yates02}, Fig. 7).
The rodent densities, in animals per hectare, are shown in log scale
to emphasize the behavior at low densities. Shown: total population
(line-dot), infected with SNV (line), critical density (dashed). The
time axis counts months from December 1994. See the text for a
discussion of the dynamics.}
\label{fig:threshold}
\end{figure}

The model of Abramson and Kenkre (henceforth mentioned as the AK
model) predicts a critical transition from a state with a positive
prevalence of the infection in the population (an \emph{infected
phase}), to a state without infection. This latter state occurs when
a parameter characterizing the environment and controlling the
population size drops below a certain threshold value, which we call
the ``critical carrying capacity.'' The population value associated
with this critical carrying capacity, $m_c$, is a population
threshold for the system: the total population, $m$, needs to be
greater than $m_c$, to be able to sustain infection at all. In the
terminology of the basic reproduction number, $R_0$, this critical
condition corresponds to $R_0=1$. Above the critical carrying
capacity, the infection is able to sustain itself (equivalent to the
condition $R_0>1$). This threshold phenomenon has been documented in
the Four Corners region, as shown in Fig.~\ref{fig:threshold}. These
population densities have been published by \citet{yates02}, where
they discuss the role of El Ni\~{n}o on the prevalence of the SNV
infection. We plot here the populations on a logarithmic scale to
emphasize the dynamics for low density values. The line with circles
shows the total \emph{P. maniculatus} density (mean density of two
nearby sites of the field study), which persists during the 7 years
of observation beginning in December 1994. The line without symbols
shows the density of mice infected with SNV. The dashed line
represents the critical population able to sustain a positive
prevalence of infection, as predicted by the AK model, by using
approximate parameters obtained from the time series (details of the
calculation can be found in \citet{abramson04}). Let us briefly
analyze the implications of this graph. At $t<15$ the population is
above the critical threshold level and, correspondingly, there is a
positive infected phase. Within this regime, it is conceivable that
the environmental carrying capacity has some time dependence, but
that it remains above the critical value. Around $t=15$, the
carrying capacity then drops to some value below criticality, and
consequently the total population $m(t)$ drops as well, approaching
an equilibrium which is now below the critical value $m_c$. The drop
is not monotonic: there are some discrete steps of decreasing
population. Concomitantly, as $m$ begins to decrease (the indication
that the carrying capacity has gone subcritical), the infected
population $m_I$ begins to disappear sporadically. A few infected
mice may be entering by migration, but it is clear that the
infection is disappearing from the site. After $t=30$ there begins a
steady population explosion, indicating that the carrying capacity
has increased, and it is observed that the population grows beyond
$m_c$, indicating that the system is again above the critical level.
The prevalence of infection is expected to recover. This process
takes time, however, just as in the AK model [see Fig. 2 in
\citep{abramson02}], and not before $t=40$ do we see a positive
$m_I$ again. After this, the population remains above critical, and
the infection persists. A brief excursion of $m$ below $m_c$ (at
$t\approx 70$) might be the beginning of a new extinction event, and
indeed $m_I$ reaches its lowest values since those at $t=40$. But
shortly after this, the time series ends and the analysis can not be
carried further. Observe that the drop in $m$ takes place in 2001, a
year that was particularly dry in the North American Southwest, as
can be seen in the precipitation data of Fig. 2
in~\citep{abramson04}.

A threshold in the animal density necessary to sustain a positive
seroprevalence has directly been observed in another Hantavirus
system, the Puumala Virus in association with the red bank vole in
Belgium \citep{escutenaire00}. It is conceivable, then, that other
processes which limit the population size would play a similar role
in the control of the infection. In real ecosystems mice share the
environment with many others species, competing for limited
resources with some, and being preyed on by others. Indeed,
competition and predation stand out as the main interspecific
relations for the species under consideration.

The effect of a predator is certainly akin to that of a competitor:
both tend to reduce the population under study. The role of
predation on the prevalence of a disease has been analyzed by
\citet{ostfeld04}, and previously by \citet{packer03}. As recognized
in the former, other phenomena besides predation (such as limited
food supply) may be the principal regulators of rodent populations.
However, under the hypothesis that predation is the main factor,
they find that the incidence of the disease decreases with an
increasing number of predators. An important peculiarity of that
model is that the number of predators is not affected by the
population of prey. As a consequence, the number of predators can be
used as a control parameter in the analysis of the dynamics. This
assumption is reasonable for generalist predators, which are
sustained by several prey species. A competitor species, on the
other hand, will necessarily be affected if the shared resources are
limited. The mathematical model, in such a situation, requires a
feedback of each of the populations into the dynamics of the other.
Various other details of the model of \citet{ostfeld04} and of
\citet{packer03} do not apply to the problem under consideration in
the present paper. Among these we might mention: the recovery from
infection, the increased mortality of infected animals, and the
increased predation of them. Certainly, the model could be tailored
for specific situations, and for Hantavirus as study of the effect
of predation will be a necessary complement to the competition
analysis that we propose in the present contribution.

Another interesting phenomenon---also relying upon the biodiversity
of an ecosystem and affecting the dynamics of zoonoses---has
received recent attention, as has its relevance to human health. Of
particular interest is the case of Lyme disease, analyzed by means
of empirically based computer simulations by \citet{buskirk95} and
by \citet{schmidt01}. A conceptual model of the phenomenon, termed
\emph{dilution effect}, was presented by \citet{ostfeld00a} and
further developed in \citep{ostfeld00b}. The dilution effect
consists in the fact that the prevalence of an infection, in the
vector, of an infectious agent (ixodid ticks and a spirochete
bacterium, respectively, in the case of Lyme disease) is reduced if
the population of their hosts contains a diversity of species. Ticks
(in their juvenile stages, which are the most relevant for the
transmission to humans) are non-specific in their feeding habits,
and therefore if only one species of possible hosts acts as a
competent reservoir for the transmission of the infectious agent,
the existence of other hosts (which are incompetent reservoirs,
being unable to transmit the bacterium to a tick feeding on them),
reduces the probability of infection of susceptible ticks. As a
result of this ``dilution,'' while the total population of ticks is
not affected by the biodiversity of their hosts, the prevalence of
the infection \emph{is} reduced. The mechanism by which this
reduction takes place in the tick population is, then, different
from the one we propose in the present contribution. The prevalence
of the infection in the Hantavirus host is reduced by the presence
of competitors at its own trophic level, the species with which the
host shares resources. These competitors exert a pressure on the
host population as a whole, and the infected sub-population may be
reduced and even led to extinction (even though the total population
persists, as we will show below). In other words, with the dilution
effect there is a reduction in the infection by means of a reduction
of the exposure, v.g. of the chance of encounter between the
infectious agent and a susceptible host. What it is shown in the
present work is a different phenomenon. Even within the host
population, the relative weight of the pathogen is reduced. We
stress the ecological importance of this feature, rather than its
mathematical justification.

With the purpose of gaining insight into the system, we aim to
understand the role of the different processes separately, while
keeping the model analytically manageable. For this reason we have
chosen to make a number of simplifications. As mentioned above, we
restrict our discussion to competition interactions only, which
arise among rodent species for a number of reasons. To mention a
few, territory, feeding habits, behaviors, etc., all contribute to
competitive behavior. In general, a particular Hantavirus has a
specific host, a single species that acts as a reservoir. This host
remains infected and maintains the infection in the population by
horizontal transmission only. It is known that, in some cases, there
may be ``alternative host species,'' that are able to host the
virus, but for some reason are unable to transmit it successfully.
Other mammals that come into contact with an infected individual of
the host species may become infected. However, non-host animals
normally represent a dead-end for the virus, which is eliminated by
the immune response of the animal~\citep{peters99}. In summary,
these species do not play a direct role in the spreading of the
disease in the wild, but they exercise, nevertheless, an ecological
pressure on the host, that eventually may affect the dynamics of the
zoonosis.

In the following section we present a generalization of the AK model
in the simplest case, incorporating just one competitive species
that cannot be infected, and studying its effect on the dynamics of
the infection. The theoretical insight obtained may suggest ways for
controlling the disease agent and, therefore, its incidence. Our
results show that competition reduces, in a specific way, the
prevalence of the infection. Interestingly, this is in agreement
with (and provides theoretical support to) a hypothesis that has
been recently put to experimental test in populations of \emph{Z.
brevicauda}, the host of the Calabazo Hantavirus, in field studies
in Panama\footnote{\raggedright Suz\'{a}n, G. (University of New
Mexico), personal communication, and also \citet{suzan04}. In their
work, only the host was left at specific sites, removing the
competing species, with the purpose of observing the change in the
infection prevalence with respect to control sites, where all the
competitors coexist.}. Similarly, also in Panama, it has been
proposed that the maintenance of competitive populations may serve
to reduce the risk to human populations exposed to \emph{O.
fulvescens} infected with the Choclo Hantavirus\footnote{Koster, F.
(Lovelace Sand\'{\i}a Health System), personal communication.}. The
proposal which has been called a ``moat'' consists of an area
surrounding human habitation maintaining a diversity of innocuous
species, competing with the hosts of the Hantavirus.

As in the AK model, we study the properties of the system as a
function of the parameters that characterize the (eventually
changing) environmental conditions. These parameters are
proportional to the carrying capacities of the system (one for each
species), and therefore provide a natural choice to analyze the
dependence of the prevalence of the infection as controlled by
limiting resources, as suggested by the trophic cascade proposed by
\citet{yates02}. The role of environmental factors that affect
rodent density, such as plant biomass and coverage, as determinants
in the seroprevalence of Hantavirus, has also been suggested by
\citet{biggs00}. The potential impact of this relationship with the
human risk of HPS has been successfully put to test by the use of
remote sensing techniques and GIS, as reported by \citet{boone00}.

\section{Intra- and interspecific competition}
We consider two species of mice, one of which is a reservoir of
Hantavirus (the \emph{host} species), and another which is not (the
\emph{alien} species). The host is subdivided into susceptible and
infected populations. Let us consider that both species interact by
competing for common resources. A usual description of this
situation, as discussed for example by \citet{may81}, is the
following model for the population dynamics of the total host, $m$,
and alien, $z$, populations:
\begin{subequations} \label{seq:bio}
\begin{align}
  \frac{dm}{dt}&= (b-c) m - \frac{m}{K} ( m + q z), \label{eq:native}\\
  \frac{dz}{dt}&= (\beta-\gamma) z - \frac{z}{\kappa} (z + \epsilon
  m), \label{eq:alien}
\end{align}
\end{subequations}
where, for the host species, $b$ is the birth rate, $c$ is the death
rate, $K$ is the carrying capacity in the absence of an alien
population ($z=0$), and $q$ is the influence of the alien
population; for the alien species, the analogous parameters are
$\beta$, $\gamma$, $\kappa$ and $\epsilon$ respectively. We consider
that the interaction is not necessarily symmetric, i.e. $q\neq
\epsilon$. One of the species can take resources from the other in a
quantity greater than what it is losing.

There are four significant equilibria for the system defined by
Eqs.~(\ref{seq:bio}), the others involve negative populations. Two
of them consist in one species surviving over the extinction of the
other. Another one is the extinct phase where all populations are
zero. Finally, there is a \emph{coexistence state} where both
species are present, and in this we will focus our interest. When
only one species persists, the equilibrium population is either:
\begin{equation}
m^*=K(b-c),
\end{equation}
or:
\begin{equation}
z^*=\kappa(\beta-\gamma),
\end{equation}
where the asterisk denotes equilibrium states. If the two species
coexist, then:
\begin{subequations}\label{coexisting}
\begin{eqnarray}
    m^*&=& \frac{K (b-c) - q \kappa (\beta-\gamma)}{1-q \epsilon}, \\
    z^*&=& \frac{\kappa(\beta-\gamma) - \epsilon K (b-c)}{1-q \epsilon}.
\end{eqnarray}
\end{subequations}
The stability of these equilibria is immediate with the use of
linear stability analysis [see, for example, \citep{murray} or
\citep{guckenheimer83}]. The conclusion of such analysis is as
follows. If the intensity of the interacting competition is not very
high, $q < 1$ and $\epsilon < 1$, then the coexistence state given
by Eq.~(\ref{coexisting}) is stable.  On the other hand, if the
competition is strong, $q>1$ and $\epsilon
> 1$, bistability occurs: the final state depends on the initial
conditions.  Finally, if $q>1$ and $\epsilon<1$ (or $q<1$ and
$\epsilon>1$), only the strong competitor survives.  For a more
complete survey see \citep{may81} or \citep{murray}.

With the same considerations made in the formulation of the AK
model, we introduce an internal classification into the host
population. The internal states consist of the infected
subpopulation, $m_{i}$, and the susceptible subpopulation, $m_{s}$.
It is clear that these states are mutually exclusive, and that their
sum recovers the total host population, $m=m_s+m_i$. The evolution
equations for these are modeled, also as in AK, according to the
following field observations: infected individuals are generated by
pair-wise interactions between susceptible and infected; there is no
vertical transmission (from infected mother to offsprings); infected
individuals die with the same rate as susceptible ones (indeed, the
virus does not appear to affect any physical or behavioral parameter
of the infected individuals). It is worth mentioning that the lack
of vertical transmission is, in the AK model, crucial for the
character of the phase transition controlled by
the---environmentally dependent---carrying capacity. If it is
relaxed, even infinitesimally, the threshold of the carrying
capacity becomes zero, and there is a positive prevalence for all
values of $K$. The resulting model is now:
\begin{subequations}
\begin{align}
  \frac{dm_s}{dt}&= b m - c m_s - \frac{m_s}{K}(m +q z) - a m_s m_i,\\
  \frac{dm_i}{dt}&= - c m_i - \frac{m_i}{K}(m+q z) + a m_s m_i, \\
  \frac{dz}{dt}&= (\beta-\gamma) z - \frac{z}{\kappa} (z + \epsilon
  m),
\end{align}
\label{eq:3species}
\end{subequations}
where $a$ is the contagion rate. The equation for $m=m_{s}+m_{i}$ is
the same as in Eq.~(\ref{eq:alien}). The AK model is recovered by
setting $q=0$, and ignoring the equation for $z$. The existence of
the two subpopulations of $m$ makes each of the equilibria
corresponding to $m\neq 0$ split into two, one of them with a
positive prevalence (an infected phase) and one without infection.
The former, for which the three populations are positive, is:
\begin{eqnarray}
    m_s^*&=& b/a, \\
    m_i^*&=& \frac{K(b-c)-q \kappa(\beta-\gamma)}{1-q\epsilon} - b/a,\\
    z^*&=& \frac{\kappa(\beta-\gamma) - \epsilon K (b-c)}{1-q\epsilon}.
\end{eqnarray}

Due to the fact that the coexistence equilibrium of
Eq.~(\ref{coexisting}) can be expressed as:
\begin{equation}
m^*= K(b-c) - q z^*,
\end{equation}
we can cast the fraction of infected mice, $\chi^{*}=m_i^*/m^*$, as
\begin{equation} \label{eq:chidez}
\chi^* = 1 - \frac{b/a}{K(b-c)-q z^*} = 1 -
\frac{K_c^{AK}}{K-\frac{q}{b-c} z^*},
\end{equation}
where $K_c^{AK}=b/a(b-c)$ is the critical carrying capacity of the
AK model, below which the infected phase disappears. This last
result coincides with the AK model when $z=0$, so that if
$K>K_c^{AK}$, and then $\chi^* >0$, the infected phase is stable,
and if $K<K_c^{AK}$ it is unstable. In the presence of the alien
population ($z>0$), we can generalize the critical parameter as:
\begin{equation}
K_c = K_c^{AK} + \frac{q}{b-c} z. \label{kc}
\end{equation}

\begin{figure}[ht]
\includegraphics[width=10cm]{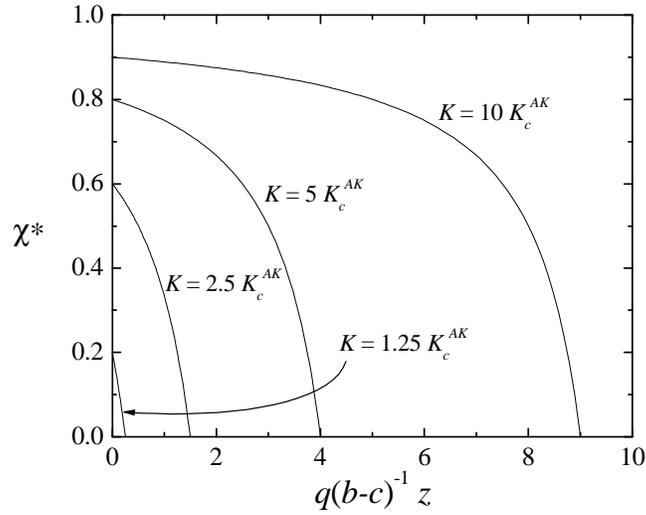}
\caption[Fraction of infected and alien population]{The fraction of
infected mice in equilibrium, $\chi^*$, as a function of the
population of aliens, $z$ (rescaled by $q/[b-c]$). We see that the
infected phase is forced to extinction by a critical value of the
alien population, that depends on the environmental carrying
capacity for the host.}
\label{fig:chiz}
\end{figure}

Let us analyze the relevance of Eqs. (\ref{eq:chidez}) and
(\ref{kc}). The former, Eq. (\ref{eq:chidez}), states that the
fraction of infected mice in the host population is reduced by the
presence of alien mice, the host's competitors.  A graphical
representation of this result is shown in Fig.~\ref{fig:chiz}, where
we plot $\chi^*$ as a function of the alien population, for several
values of the carrying capacity $K$. Fig.~\ref{fig:chiz} contains
one of the fundamental results of the present work, clearly showing
that the competitive pressure exerted by the alien population has
the effect of reducing the prevalence of infection in the host. In
fact, from Eq~(\ref{eq:chidez}) it can be seen that the prevalence
is maximum when no alien population is present, $\chi^*(z>0) <
\chi^*(z=0)$, and that increasing the amount of aliens will always
reduce the prevalence, since $\partial \chi /\partial z<0$. A
further important result of the present calculation is the existence
of a critical amount of aliens, a threshold level in the population
of competitors, that drives the system completely to a non-infected
state. This value is given by
\begin{equation} \label{zc}
z_c = \frac{a K(b-c)-b}{a q}.
\end{equation}
When the environmental parameter $K$ has a value greater than the
critical $K_c$ the system has a positive prevalence of infection. On
the other hand, when $K<K_c$, the competitor population results
greater than the minimum necessary to force the infected
subpopulation to extinction. The point $K=K_c$ constitutes a
critical point for the system, separating two behaviors that
qualitatively differ in the stability of the equilibrium of the
infected population. Observe that time-varying environmental
conditions, as well as a heterogeneous landscape, may provide a
framework where a wealth of spatio-temporal patterns of infection
would occur.

\begin{figure}[ht]
\includegraphics[width=13cm]{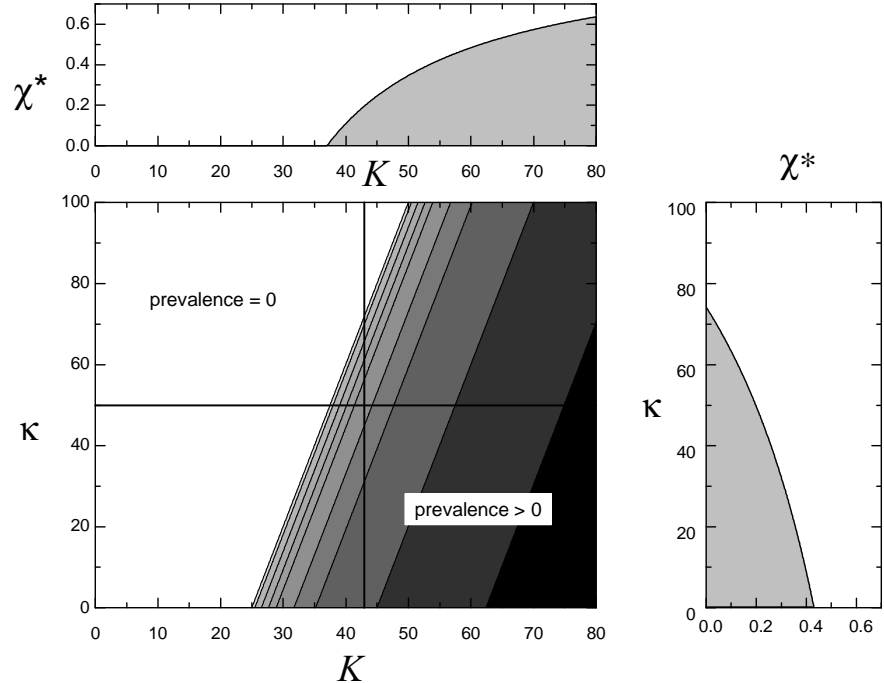}
\caption[Phase diagram]{Phase diagram of the infection prevalence,
in the space defined by the host carrying capacity $K$ and the alien
carrying capacity $\kappa$. The contour plot shows the prevalence as
shades of grey, with the darkest shades corresponding to higher
prevalence. The white region is the region of zero infection. The
crossing lines indicate sections of the plot, shown in the upper and
right side plots, where the prevalence is shown as a function of the
relevant control parameter. The right panel represents the
extinction of the infection controlled by the alien carrying
capacity. The parameters, in this example, are: $b=1$, $c=0.6$,
$q=0.2$, $\beta=1$, $\gamma=0.5$, $\epsilon=0.1$, $a=0.1$.}
\label{fig:phases}
\end{figure}

In Fig.~\ref{fig:phases} we show a phase diagram that provides a
complementary picture to that given by Fig.~\ref{fig:chiz}. The
shaded contour plot shows the prevalence (as scales of grey) in the
parameter space defined by the two carrying capacities, $K$ and
$\kappa$. The region painted white, that includes the origin, is the
region free of infection, where the state with $m_i^* = 0$ is
stable. The first contour defines the transition line, beyond which
a positive and stable prevalence is to be found, in a potentially
infinite region in parameter space. Two sections of the phase plot
are also shown, along the lines perpendicular to the axes seen in
the contour plot. Each one of the sections represents a projection
of the prevalence along one of the carrying capacities. Both of them
are in the form of a transcritical bifurcation, in which two
equilibria of different character exchange their stability. The
section shown at the top of the figure represents the transition as
a function of the carrying capacity of the host, $K$. There is a
threshold above which the population shows a positive prevalence. A
similar picture characterized the AK model (as a function of $K$
only) [see Fig. 1 in \citep{abramson02}]. The presence of the
competitor reduces the seroprevalence in such a way that the
bifurcation is not a linear relation between the equilibrium
solution and the control parameter. The section shown on the right
panel represents the effect of the carrying capacity of the
competitor, $\kappa$. In this case, a greater carrying capacity
inhibits the infection, because it allows a greater population of
competitors.

It is important to remark that the ``strength'' of the competition
with the alien population, $q$, is the same for both susceptible and
infected hosts [Eq.~(\ref{eq:3species})]. Indeed, both
subpopulations are reduced as a result of the competition, but the
infected population suffers the consequences in a stronger manner,
becoming extinct in a critical way. This phenomenon, analogous to a
second order phase transition in a physical system, occurs at a
finite value of the density of alien mice, $z_c$ as shown in
Eq.~\eqref{zc}.

The origin of the differential effect by which susceptible and
infected hosts feel the competitive pressure differently, deserves
some discussion. This effect is caused by an important difference
between susceptible and infected mice that---even though it was
already mentioned when the classification was introduced---until now
had its true relevance not exposed. The difference resides in the
ways in which new members of each type are incorporated in the
population. While the susceptible are born, the infected appear only
by contagion of already born (and susceptible) mice. In biological
terms, there is no vertical transmission, a fact that produces a
strong asymmetry between susceptible and infected. The lack of a
differential death rate between susceptible and infected animals is
also a peculiarity of this system, which is uncommon in other
epizootics. These two ingredients, a difference and a similitude
between the two classes of host animals, are responsible for the
differential effect.

\section{Conclusions}
We have studied a model of the epizootic of Hantavirus in a system
composed of two rodent populations: a host of the virus, which
remains chronically infected by horizontal transmission, and a
competitor species that does not host the virus. This situation is
paradigmatic of many Hantavirus-rodent systems, in which a single
species hosts the virus in a competent way, and several other
species compete with the host. The model generalizes a previous
model by \citet{abramson02} [also \citep{abramson03}], that
described the dynamics of the host only. The interaction between the
two rodent species, since it is a competition for limited resources,
produces a mutual limitation of the populations. In particular, the
second species exerts a regulatory effect on the host. This
competitive pressure is the same for both susceptible and infected
hosts (as it should, since the infection with Hantavirus does not
affect any biological parameters in the hosts). Nevertheless, a
reduction of the seroprevalence in the host species is observed, up
to a point where it completely disappears from the system. This
phenomenon can be described as a critical transition between an
infected phase and a phase free of infection, controlled by
the---environmentally determined---carrying capacities of the two
species. In the space defined by these carrying capacities spatially
or temporally varying environmental conditions would be described by
a path, which might cross the transition line, thus driving a
corresponding change in infection phase. This is analogous to the
results discussed by \citet{abramson02} and \citet{abramson03}, as
well as, conceptually, by \citet{yates02}. Furthermore it is
interesting  that a change \emph{in the carrying capacity of the
competitor only} (a vertical path in the phase space of Fig.
\ref{fig:phases}) is enough to drive the infection to extinction. In
a real world situation, where the carrying capacities exhibit
temporal as well as spatial heterogeneity, a wealth of dynamical
transitions are to be expected.

Alternatively, the competitor's population can be used to define an
effective carrying capacity of the host, which exhibits the same
threshold behavior. Our model has been intentionally kept simple to
allow a direct analysis of the consequences by analytical means.
This thus provides a manageable framework to study a necessary step
towards the modeling of realistic systems, which certainly contain
several species of rodents and small mammals in competitive
interaction, as well as other species in the trophic web. The study
of similar systems, for example: when competing species are
organized in a hierarchical way \citep{tilman94}; the role of
fluctuations when populations are small; etc., are currently under
study and will be presented in forthcoming contributions.

\acknowledgments{The authors acknowledge fruitful discussions with
Gerardo Suz\'{a}n and Fred Koster, as well as the valuable comments made
by the referees and the handling editor of the manuscript. We are
also grateful to Maureen Ballard for her careful reading of the
manuscript and her many suggestions. G. Abramson acknowledges
financial support  from CONICET (PEI 6482) and ANPCyT (PICT-R
2002-87/2). Further support has been provided by the Consortium of
the Americas for Interdisciplinary Science (University of New
Mexico) through the following grants: INT-0336343 (NSF's
International Division) and DARPA-N00014-03-1-0900. As always, it is
a pleasure to acknowledge the hospitality of the Abdus Salam ICTP
(Trieste). }

\end{document}